\documentclass[hidelinks,manuscript]{elsarticle}

\usepackage[a4paper, margin=38mm]{geometry}

\usepackage{lineno,hyperref}
\modulolinenumbers[10]

\let\linenumbers\nolinenumbers\nolinenumbers

\journal{METHODS}

\usepackage{setspace}
\usepackage{calc}
\usepackage{xspace}
\usepackage{amsmath}
\usepackage[normalem]{ulem}

\usepackage[printonlyused]{acronym}

\newcommand{\prog}[1]{\mbox{\emph{#1}}}

\newcommand{\eg}{e.\,g.,\xspace}
\newcommand{\ie}{i.\,e.,\xspace}

\usepackage{xcolor}

\begin{document}
\begin{frontmatter}

\title{Evolving methods for rational \textit{de novo} design of functional RNA molecules}
\tnotetext[issue]{Issue title: Chemical Biology of RNA\\Guest Editor: Michael Ryckelynck}


\author[bioinf]{Stefan Hammer\corref{mycorrespondingauthor}}
\cortext[mycorrespondingauthor]{Corresponding author}
\ead{jango@bioinf.uni-leipzig.de}
\ead[url]{www.bioinf.uni-leipzig.de}

\author[biochemie]{Christian G\"unzel}
\author[biochemie]{Mario M\"orl}
\author[bioinf]{Sven Findei{\ss}\corref{mycorrespondingauthor}}
\ead{sven@bioinf.uni-leipzig.de}

\address[bioinf]{Bioinformatics,
Institute of Computer Science, and Interdisciplinary Center for
Bioinformatics, Leipzig University, H\"artelstra{\ss}e 16-18, 04107 Leipzig, Germany}
\address[biochemie]{Institute for Biochemistry, Leipzig University, Br\"uderstra{\ss}e 34, 04103 Leipzig, Germany}

\begin{abstract}
 Artificial RNA molecules with novel functionality have many
 applications in synthetic biology, pharmacy and white biotechnology.
 The \textit{de novo} design of such devices using computational
 methods and prediction tools is a resource-efficient alternative to
 experimental screening and selection pipelines. In this review, we
 describe methods common to many such computational approaches,
 thoroughly dissect these methods and highlight open questions for the
 individual steps. Initially, it is essential to investigate the
 biological target system, the regulatory mechanism that will be
 exploited, as well as the desired components in order to define
 design objectives. Subsequent computational design is needed to
 combine the selected components and to obtain novel functionality.
 This process can usually be split into constrained sequence sampling,
 the formulation of an optimization problem and an \textit{in silico}
 analysis to narrow down the number of candidates with respect to
 secondary goals. Finally, experimental analysis is important to check
 whether the defined design objectives are indeed met in the target
 environment and detailed characterization experiments should be
 performed to improve the mechanistic models and detect missing design
 requirements.
\end{abstract}

\begin{keyword}
RNA design \sep rational \textit{de novo} design \sep synthetic biology \sep artificial RNA devices \sep mechanistic models \sep sequence sampling \sep experimental validation \sep RNA design tools

Published at \textit{METHODS}, DOI: \href{https://doi.org/10.1016/j.ymeth.2019.04.022}{10.1016/j.ymeth.2019.04.022}
\end{keyword}

\end{frontmatter}

\linenumbers

\ac{RNA} sequences are the perfect building blocks to reprogram
cellular behavior, as they are known to regulate gene expression at
almost every step and in all domains of life. Features like
environment-sensing abilities, enzymatic reactivities and a cost
effective \textit{in vitro} or \textit{in vivo} synthesis make
\ac{RNA} a Swiss army knife in synthetic biology and its related
fields such as (white) biotechnology and personalized medicine.
Example applications include self-assembling RNA scaffolds either designed
to increase metabolic production by co-localizing related enzymes \cite{Delebecque:2012, Sachdeva:2014}
or to form lattices and tubular structures mimicking cytoskeletal proteins
\cite{Nasalean:2006, Stewart:2017}. Artificial \ac{RNA}
molecules can also be made to be highly stable by incorporating
modified nucleic acid analogs, which circumvent rapid degradation by
RNase enzymes \cite{Khvorova:2017}. Early clinical trials utilizing
such molecules as drugs show that \ac{RNA}-based therapeutics might be
an alternative route to cure so far untreatable (genetic) diseases
\cite{lieberman_tapping_2018}. However, these medical applications are
still in their infancy and challenges, such as intracellular delivery
across membranes to target specific tissues, need to be tackled before
\ac{RNA}-based personalized medicine can become a standard approach,
as recently reviewed by \citet{lieberman_tapping_2018}. Naturally
occurring \ac{RNA} regulators and the regulatory mechanisms they
employ --- for instance miRNAs, riboswitches, trans-activating
\acp{RNA} and ribozymes --- often serve as templates for designing
artificial counterparts or to fill specific gaps in the available
repertoire of \ac{RNA} devices. The diverse set of designed functional
mechanisms was extensively reviewed \cite{qi_versatile_2014,
chappell_renaissance_2015, karagiannis_rna-based_2016,
findeis_design_2017, lieberman_tapping_2018, lee_design_2018,
Etzel:2017, ameruoso_brave_2019} and thus is not discussed here.
Advances in experimental technologies, such as high-throughput
techniques to investigate functional variants or to determine \ac{RNA}
structure, and in computational biology make \ac{RNA} design a growing
and fast developing research field.

As summarized by various reviews \cite{mckeague_opportunities_2016,
jang_rna-based_2018, sherman_computational_2018}, artificial \ac{RNA}
molecules can be generated by several experimental strategies, such as
selection and screening approaches, and by computational rational
design. In this contribution, we will focus on how rational \textit{de
novo} design of single- and multi-stable \ac{RNA} molecules was
accomplished in the literature. Hence, computational methods and
experimental strategies to investigate the designs including their
essential interplay are summarized, but we generally disregard the
actual mechanisms of the novel \acp{RNA} in this review.

\begin{figure}[!ht]
  \centering
  \includegraphics[width=0.9\textwidth]{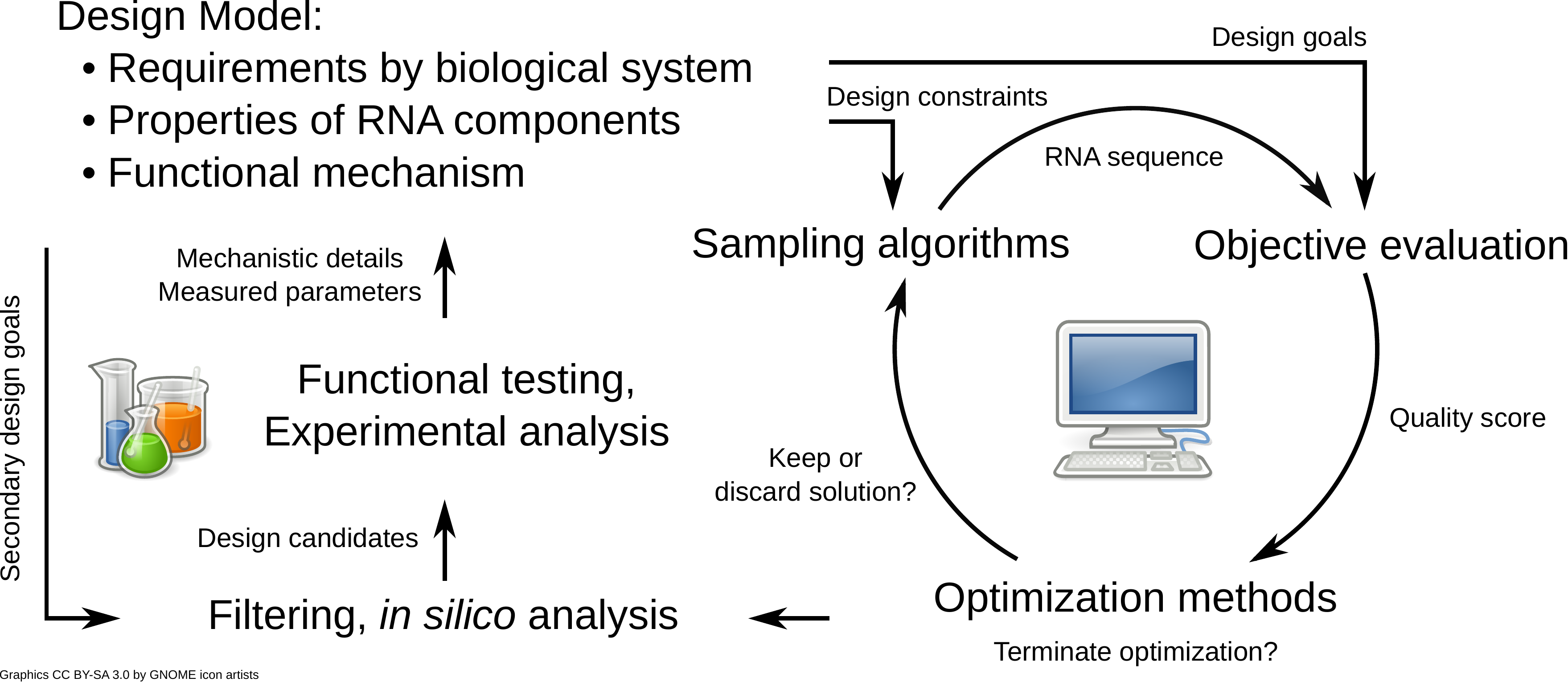}
  \caption{Overview of a general \textit{de novo} design process. The
initial design model development comprises a detailed investigation of
the biological testing system, how to interface with the target
environment, compile properties and parameters of \ac{RNA} components
to be utilized and drafting the functional mechanism of the overall
design. Computational design is necessary to connect components while
maintaining their individual functionality and to gain the desired new capabilities. Software tools require design constraints and design
goals, which are formulated from the model, as input. An iterative
optimization loop of sequence sampling, objective evaluation and
decision-making is performed. A subsequent filtering and \textit{in
silico} analysis step helps to include secondary design goals,
properties which could not be taken care of in the optimization
procedure. Functional testing and detailed experimental analysis in
the lab are intended to identify working devices and to deliver
valuable details and measured parameters to improve the design model.}
  \label{fig:pipelineoverview}
\end{figure}

We previously argued~\cite{Hammer:2018} that attempts to rationally
\textit{de novo} design \ac{RNA} molecules frequently follow a common
construction that we view as a pipeline, see \autoref{fig:pipelineoverview}. Approaches to design RNAs
consistently include initial analysis of the biological target system
and the desired components followed by the application of one or more
computational methods to derive candidate \ac{RNA} sequences. Many
design approaches also include further \textit{in silico} analysis to
reduce the list of candidates that must be tested and characterized in
the laboratory. We found many similarities within the various
design studies, for example, which algorithms and computational
methods were used as well as the experimental setup applied for
validation and analysis. We therefore decided to review the three
pipeline steps individually in the subsequent sections, describe how
they were accomplished and highlight state of the art methods and
novel ideas.

\section{Characterization of the utilized biological system and
 components}
\label{sec:characterization}

Any rational \textit{de novo} approach needs \textit{a priori}
knowledge about the relevant biological systems and components in
order to reliably make decisions that finally lead to functional
designs. Missing knowledge about the biological system and components,
or premature design decisions will cause an extended experimental
testing phase or might even make the project unfeasible.
Unfortunately, this phase of information gathering is rarely described
in detail. We therefore devote this section to summarizing what kind
of information needs to be collected for a successful RNA design
project.

The essential first step of designing functional \ac{RNA} molecules is
a detailed characterization of the \textit{in vitro} or \textit{in
vivo} system in which the \ac{RNA} must function. Typically, the novel
\ac{RNA} molecule is designed to interact with its environment in a
specific manner. However, it might also influence --- and be
influenced by --- other factors. Thus, it is not only important to
examine the mechanism and properties of the desired interaction
between the RNA device and its environment, but even more crucial to
consider unwanted secondary interactions. Examples therefore might be
off-target binding of the artificially designed RNA, interactions with
proteins or other molecules, or degradation by RNases. If known, such
disruptive factors can be taken into account at the various design
steps. To decrease such influences during testing and debugging
experiments, \textit{in vitro} approaches such as the PURE cell-free
system \cite{Shimizu.2005} and microfluidics devices
\cite{Autour:2017, Ryckelynck_Microfluidics_2015} can be used to
approximate \textit{in vivo} conditions while utilizing a well-defined
reaction environment.

Moreover, researchers face many design decisions, from whether to use
specific reporter genes or vector systems, to the selection of
auxotrophic markers or resistance genes, and all these decisions impact
the cell, the constructed plasmid and the functionality of the novel
\ac{RNA} device. For example, many strains exist for common model
species, such as \ac{ecoli} strains like TOP10, DH5$\alpha$ or MG1655.
They exhibit distinct features which affect the applicability of
vectors or detection methods. For instance, TOP10 cells --- in
contrast to MG1655 or DH5$\alpha$ --- combine the inability to
metabolize arabinose with the deletion of the \textit{lacZ} gene.
Thus, TOP10 cells allow for systems that use arabinose-inducible
\textit{araBAD}-promoters and $\beta$-galactosidase as a reporter
system. Therefore, it is important to carefully consider such issues
about the biological testing system before any computation, as these
early decisions affect subsequent steps in the project. For example,
the sequence context of the vector or the reporter gene sequence must
be taken into account, either directly --- as input for design
programs --- or indirectly --- as secondary design goals. The latter
could be seen as good-to-have goals, where possible design candidates
are sorted and filtered by their predicted \ac{RNA}-\ac{RNA}
interaction with the context, in this example. Moreover, early
decisions such as bacterial strains constrain what type of experimental
analysis and characterization is possible, as explained in more detail
in \autoref{sec:expanalysis}.

Another example of a parameter which needs to be included as a design
goal for the \textit{in silico} computations are \ac{RNA}
concentrations. The relative amount of \acp{RNA} present in the cell
is critical for proper functionality. This issue is especially
crucial when an artificially introduced RNA must interact with other
molecules, \eg a riboswitch that is induced by a small RNA. RNA
concentrations can be specified by using strong or weak promoters,
whose strength is determined by their affinity for a specific
$\sigma$-factor or \ac{RNA} polymerase. Additionally, gene expression
levels are also affected by the copy number of the chosen vector,
which is regulated by the origin of replication (\eg high copy ORI
pMB1 vs. low copy ORI p15A). When multiple plasmids are used, their
origins of replication need to be compatible, as cloning procedures,
\eg for special resistance mechanisms, otherwise become difficult.

Additionally, during the design phase, information needs to be
collected about biological components which will be part of the novel
\ac{RNA} device such as promoter, aptamer, catalytic centers, \ac{RBS}
or \ac{SD} sequences. The iGEM Registry of Standard Biological Parts
is a web resource\footnote{\url{http://parts.igem.org/}} which
collects well-specified standard and interchangeable components
that can be used to design and construct integrated biological
systems. Detailed information on these components is crucial for
computational design. Such properties include structural
conformations, essential nucleotides and the reaction mechanism (\eg
activity levels, reaction rates, binding affinities, ligand chemistry
or reaction intermediates). The sequence context of a component is
also important, and must be included as component property. For
example, a stretch of consecutive uracil nucleotides in front of a
\ac{SD} sequence can boost translation initiation~\cite{Zhang:1992},
while stable structural elements can cause the opposite
effect~\cite{Borujeni:2013}.

\begin{figure}[t]
  \centering
  \includegraphics[width=0.5\textwidth]{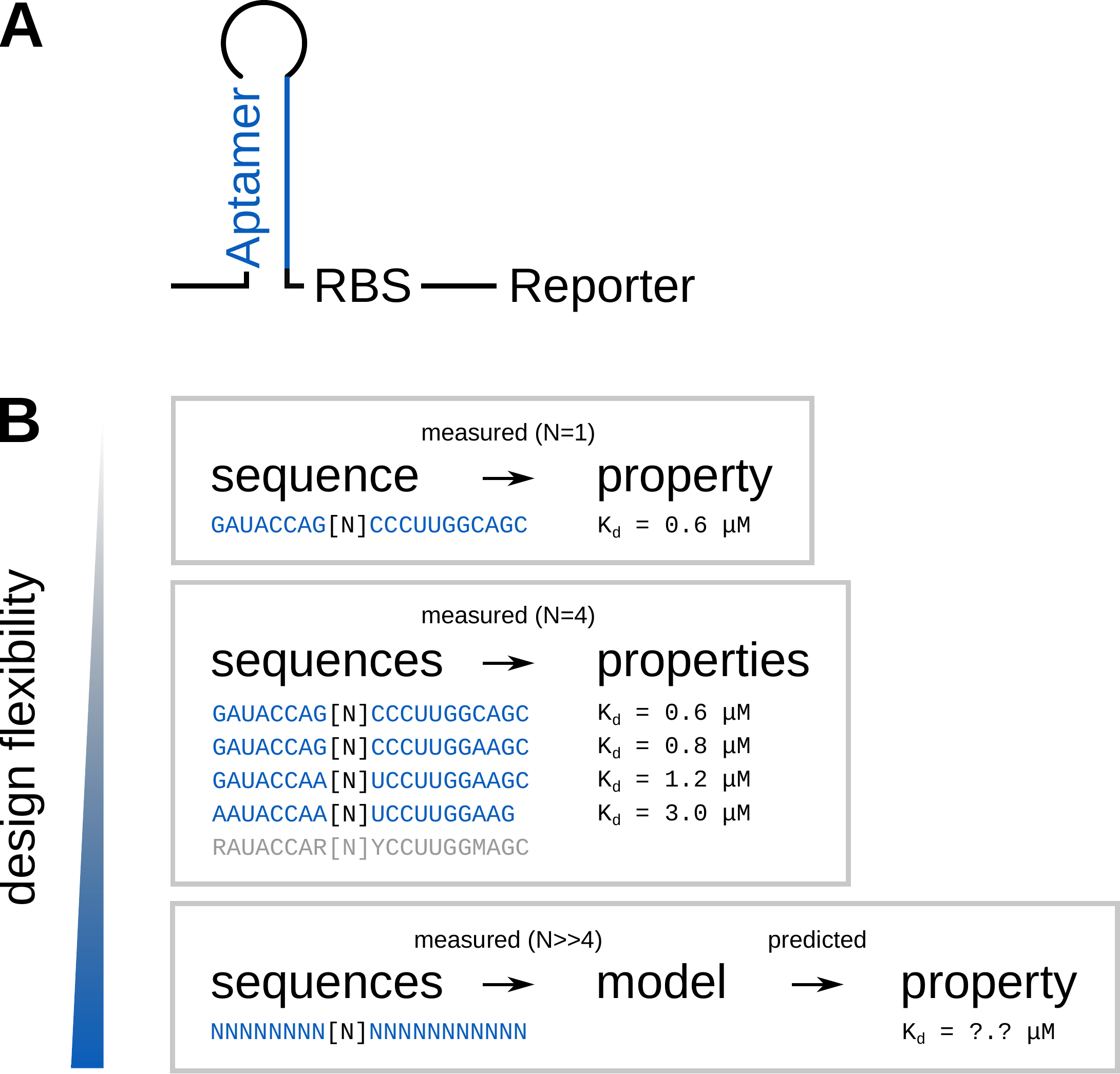}
  \caption{\textbf{A)} Artificial \ac{RNA} devices with novel
functionality are usually constructed by combining various components,
\eg an aptamer, a \acf{RBS} and a reporter gene. \textbf{B)} The
various properties of these components, and thus the functionality,
need to be provided to computational models. Directly measured
properties of a single sequence (top) --- in this example the binding affinity of
the aptamer --- fulfills this task, however with no design flexibility
for that component. A catalog of various sequences increases the
flexibility (middle), while a dedicated model --- constructed or
trained by a big dataset --- allows for maximum flexibility, as the
properties can be predicted for any arbitrary sequence.}
  \label{fig:component}
\end{figure}

In addition to extracting this information from the literature, it is
important to perform experiments to characterize individual components
with respect to the selected target system, in order to verify their
functionality in the novel environment. For example, an aptamer needs
its ligand binding affinity measured and the binding structure
confirmed in conditions similar to the target environment. These measurements deliver the valuable
sequence to function relations, which are crucial parameters for
computational models (\autoref{fig:component}).

Many studies created \ac{RNA} designs by using fixed
nucleotide sequences as \ac{RNA} components. Random linker regions were then inserted to derive novel functions. For example, a
fixed aptamer sequence with its known binding
affinity~\cite{wachsmuth_novo_2013, Domin:2016} or an \ac{RBS}
sequence and its translation efficiency~\cite{Green:2014}.
Although this approach provides a perfect sequence to function mapping
for the individual components, it leads to very restrained flexibility
for the computational design algorithms and thus to \ac{RNA} devices
with presumably bad overall functionality. For more flexibility,
components should be described with as few fixed nucleotide positions
as possible. To increase the number of possible sequences for a
component, while still employing a directly measured sequence to
function relation, a catalog of various sequences for the component
and their corresponding function could be used. The \textit{in silico}
design algorithm could then select a component sequence which fits the
overall \ac{RNA} device best (\autoref{fig:component}).

Ideally, the desired components are described by an individual
mechanistic model that maps any nucleotide sequence to a functionality
score and \textit{vice versa}, given user-supplied environmental
factors such as temperature (\autoref{fig:component}B). These models could be derived
through high-throughput experiments and statistical learning or
through extensive experimental characterization. Such models have the
big advantage, that we might be able to predict the function for any
given sequence, even for those which were not directly measured. An
impressive example of a successful design approach based on detailed
mechanistic models is the riboswitch design by \citet{Borujeni:2015}.
The underlying model to estimate translation initiation
efficiency has been developed over years and is reported in several
successive publications \cite{Salis:2009, Salis:2011, Borujeni:2013,
Borujeni:2016}. Based on this mechanistic model, they designed riboswitches and
accurately predicted their overall functionality.
 
Yet, how much flexibility is indeed necessary to obtain good design
results depends a lot on the complexity of the desired functional
mechanism and on the degree of novelty. The latter ranges from small
performance improvements of an existing device, where solutions are
likely to be highly similar to a previously established device, to a
device with novel functionality designed completely from scratch.

\section{Computational design of \ac{RNA} molecules}
\label{sec:computationaldesign}

After determining all the details of the target system and the set of
components involved, it is necessary to use computational methods.
This design step connects the selected building blocks and adds
additional functionality, \eg ensuring that the desired regulation
will take place. All of this computational analysis must also take the
target biological environment into account. 

Many previous attempts disregarded this step and the designs were assembled by solely combining fixed sequences for
the desired components \cite{isaacs_engineered_2004,
neupert_design_2008, dawid_rna_2009, wachsmuth_novo_2013, Domin:2016}
and inserting variable regions or changing various nucleotides to
achieve the desired functionality. This approach is
prone to the generation of dysfunctional devices, as even single
nucleotide mutations can greatly alter the structure
\cite{schuster_sequences_1994} and thus disrupt the functionality of
the components. Therefore, after these alterations are introduced,
computational prediction is needed to ensure that components still
work in the new sequence context and that they perform as specified by the
design goals. 

Alternatives to a computational design approach are \textit{in vitro}
and \textit{in vivo} screening or selection pipelines
\cite{mckeague_opportunities_2016}, where specific component sequences
are concatenated with additional variable or random regions, leading
to a mixed pool of different candidates \cite{weigand_screening_2008,
lynch_throughput_2007, harbaugh_screening_2018}. Experimental
measurement of the overall functionality of the individual devices
leads to the elucidation of well-performing candidates. However, these
approaches do not scale well with the complexity of the design
mechanisms. Early design decisions, such as where random regions are
placed, which exact component sequences are chosen or where sequence
variability is allowed, will inevitably limit the solution space.
Thus, when bad initial decisions are made, the likelihood of finding
optimal devices is limited or --- for complex mechanisms --- can be
vanishingly small. Furthermore, these screening or selection pipelines
only allow for the measurement of the overall functionality coupled to a
readout mechanism, and not the functionality of the individual
components. If mechanistic insights are desired, additional
experimental analysis as outlined in section~\ref{sec:expanalysis} is
needed. Moreover, in the absence of a high-throughput readout for the
desired functionality, a screening or selection requires extensive
laboratory effort.

In contrast, computational algorithms are capable of testing thousands
to billions of sequence combinations. The predictive power of models allows to evaluate the functionality of any arbitrary sequence, in order to find an optimal overall solution in a
huge solution space. The capabilities of computational approaches are,
of course, limited by the accuracy of the underlying models and
parameters. For instance, structure prediction is only as good as the
underlying experimentally measured energy parameters and the actual
fold of a molecule might depend on environmental parameters that are
not included in the model. Thus, although computational approaches are
powerful tools to estimate \ac{RNA} device performance, the
predictions certainly need to be verified experimentally.

\subsection{Design goals}
\label{sub:designgoals}

Rational design starts by using the previously collected set of
requirements, parameters and functional mechanisms to define design
objectives which can be understood by design software and prediction
algorithms. Many design programs focus on secondary structures as
their main input, as an \ac{RNA}'s secondary structure is closely
related to its function. This simplification facilitated the
development of many design programs, as it allowed complicated
conditions to be broken into two main design goals, called
\textit{positive design} and \textit{negative design}.
\textit{Positive design} means that the target structure must be
thermodynamically stable, while \textit{negative design} ensures that
contrary structures are less stable and thus less probable in the
ensemble of structures. A sequence which folds into the desired target
structure is found, if both conditions are fulfilled.

Early design programs, such as \prog{RNAinverse}
\cite{hofacker_fast_1994}, incorporate secondary structure or features
based on the secondary structure as the major design goal
\cite{churkin_design_2017}. Designing functional \acp{RNA} is
therefore often reduced to solving the so-called \textit{inverse
folding problem}, which is the problem of finding a sequence that
adopts a given secondary structure as its most stable, \ie \ac{MFE},
state. Many programs addressed this by searching for sequences which
minimize the distance between the target and the actual \ac{MFE}
structure \cite{hofacker_fast_1994, andronescu_new_2004,
busch_info-rnafast_2006, gao_inverse_2010} or the distance to the
desired shape \cite{avihoo_rnaexinv:_2011}. \citet{zadeh_nucleic_2011}
called this general approach \textit{MFE defect} optimization and
emphasized its limitation: the probability of the target structure,
which is proportional to its Boltzmann factor~\cite{Hofacker:2007},
can still be small, due to alternative structures with similar
stability in the ensemble of structural conformations. To circumvent
this effect, the probability of the target structure in the ensemble
can be directly computed and maximized \cite{hofacker_fast_1994,
taneda_modena:_2010}. Unfortunately, optimizing for this so-called
\textit{probability defect} aims to remove any structure in the
ensemble that is not the exact specified target structure. This way, even
very similar --- and thus probably desired neighboring structures ---
are removed. In contrast, the \textit{ensemble defect}, which was
implemented as quality measurement in \prog{NUPACK:Design}
\cite{zadeh_nupack:_2011}, ensures that structures similar to the
target are favored, while distant or contrary structures are
prohibited. Even though the latter is a perfect measure to derive robust
single-stable \ac{RNA} molecules, specifying additional goals might be
helpful. For example, \textit{mutational robustness}
\cite{avihoo_rnaexinv:_2011} aims to retain the desired target
conformation and thus the functionality despite sequence mutations.
Such design goals are perfectly suited to built static \ac{RNA}
molecules like \acp{tRNA} which should reliably fold into a single
conformation.

However, many applications require that the designed \acp{RNA} are
able to change their conformation upon an external signal.
\citet{wolfe_constrained_2017} introduced a reaction pathway approach,
where the \textit{ensemble defect} is minimized for multiple \acp{RNA}
in a reaction well \cite{wolfe_sequence_2015}. The addition of trigger
\ac{RNA} molecules induces structural changes, leading to a multi-state
system by multiple \ac{RNA} strands. Designing a single \ac{RNA}
molecule which adopts multiple structural conformations is also
possible, but requires more complicated objective terms, as the
ratio and the conversion between the states has to be specified. A
bi-state objective was introduced by \citet{flamm_design_2001} in the
\prog{switch.pl} program and extended to multi-state by
\citet{honer_zu_siederdissen_computational_2013} and
\citet{hammer_rnablueprint_2017}. It maximizes the probabilities of
the target structures in the ensemble while keeping the free energies
of the target states similar. Structural change upon an external
trigger such as temperature or the addition of ligands can be
achieved by including the external factor in the various terms of one
target state \cite{flamm_design_2001, lorenz_rna_2016}.
\citet{flamm_design_2001} proposed objectives that also design the
conformation landscape by defining energy barriers between states or
the energy difference of target states.\medskip

Recent studies refrain from solving the \textit{inverse folding
problem} as the only design objective and thus relying solely on the
close structure-to-function relationship of \ac{RNA}. They rather
specify the characteristics of their desired mechanistic model
directly by including sequence-to-function measurements, crucial
properties of the used components and the desired functional mechanism
\cite{rodrigo_ribomaker:_2014, salis_automated_2009,
hu_generating_2015, hu_engineering_2018, findeis_silico_methods_2018}.
These models might still include structural features required at
specific states, but should encompass much more than structure
prediction at the thermodynamic equilibrium.
\citet{pyle_challenges_2016} as well as \citet{carlson_elements_2018}
suggest that the process of co-transcriptional folding and other
kinetically driven events such as temperature change or ligand
interaction should be incorporated directly into the design process.
Also 3D structural motifs or 3D structure prediction could be
beneficial due to the closer function relationship. A kinetic model
for co-transcriptional folding and \ac{RNA}-ligand interaction was
recently included in the \prog{ViennaRNA} program \prog{barriers}
\cite{kuhnl_tractable_2017, wolfinger_efficient_2018} and could be
used to predict this important design goal. However, a more detailed
model usually comes with higher computational costs and kinetic
simulations for thousands of candidate sequences are rarely feasible
as a design objective. Therefore, computationally fast thermodynamic
features dominate current design objectives and more detailed analyses
are usually postponed to an \textit{in silico} analysis and
verification step.

In summary, when formulating design objectives, two conditions are
crucial. The main goals should explicitly ensure that the
functionality of the individual components is not disrupted when
combined into a novel \ac{RNA} device and the overall intended
functionality should be specified as directly as possible.\medskip

\subsection{Generate a sequence towards design objectives}

To obtain a sequence which fulfills the desired goals, various
techniques were applied, including stochastic optimization approaches,
constraint programming \cite{garcia-martin_rnaifold:_2012,
garcia-martin_rnaifold_2015} or lately also machine learning
\cite{eastman_solving_2018, groher_tuning_2018}. Some of them were
recently reviewed by \citet{churkin_design_2017}. Basically, all
approaches have in common that they require a method for sequence
generation. As the solution space for \ac{RNA} sequences is
exponentially big, constraints or weights are applied to exclude
solutions that do not have essential properties. Subsequently, the
quality of the obtained sequence candidates is evaluated according to
previously specified objectives and a strategy to decide whether to
keep or discard a solution is defined. This strategy might be an
advanced optimization algorithm and/or simple rules on how to sort and
filter candidate sequences with respect to certain threshold
boundaries.

\subsubsection{Generating sequences with given constraints and features}
\label{sub:sequencegen}
Unconstrained sequence sampling, where one out of four nucleotides is
chosen per position, leads to a search space which is exponentially
growing with the sequence length ($4^n$, were $n$ is the sequence
length). Thus, sequence constraints are usually introduced to exclude
parts of the sequence space. These can be hard constraints such as
specific nucleotide patterns at fixed positions, \eg for binding
motifs and transcription start sites, as well as soft constraints
specifying nucleotide compositions like GC-content or a coding
sequence for specific amino acids. Sampling towards a probabilistic
sequence model could be also applied, \eg to obtain variable binding
motifs. Negative constraints which aim to avoid sub-sequences, \eg specific restriction cut sites, in the
complete design are also desirable.

Furthermore, introducing structural constraints is highly beneficial,
as they tend to decrease the search space enormously
\cite[Fig.\,2]{honer_zu_siederdissen_computational_2013}. For example,
ribozymes, terminators or aptamers need to adopt a specific secondary
structure to be functional. Thus, nucleotides in such a region must be
able to form the correct base-pairing pattern. However, due to the
complexity of the underlying problem, we need to distinguish between
the design of single-stable and bi- or multi-stable \ac{RNA} molecules.
Satisfying a single secondary structure constraint is easy,
as interacting nucleotides can be simply picked from the set of allowed base-pairs
at the corresponding positions. In contrast, respecting multiple
structure constraints, maybe even in combination with certain sequence
constraints, is computationally complex \cite{flamm_design_2001,
hammer_fixed-parameter_2018}. It is worth noting that crossing
base-pairs such as in pseudo-knots can be easily handled as a
structural constraint \cite{taneda_multi-objective_2015,
hammer_rnablueprint_2017}.

The first program that could uniformly sample sequences able to adopt
two structural conformations was \prog{switch.pl}
\cite{flamm_design_2001}. The corresponding contribution introduced many definitions and problem statements which are still an important foundation. Successive programs ---
\prog{multiSrch}~\cite{ramlan_design_2011},
\prog{Frnakenstein}~\cite{lyngso_frnakenstein_2012} and \prog{MODENA
2.0}~\cite{taneda_multi-objective_2015} --- then implemented decision
tree based enumeration methods which enabled to include more than two
structure constraints. However, it was shown that such methods
introduce undesirable sequence biases
\cite{hammer_rnablueprint_2017}. Therefore,
\citet{honer_zu_siederdissen_computational_2013} introduced a
graph-coloring counting algorithm --- implemented in
\prog{RNAblueprint} \cite{hammer_rnablueprint_2017} --- which allowed
to uniformly draw sequences with respect to multiple structural
constraints. Recently, it has been proven that this problem is
\#P-hard \cite{hammer_fixed-parameter_2018}.

As a compatible sequence can -- but does not necessarily -- fold into the
given target structure(s), it is beneficial to include \ac{RNA} energy
parameters in the sampling procedure and thereby obtain sequences that
are more likely to form the desired base-pairs in their structural
ensemble \cite{andronescu_new_2004, busch_info-rnafast_2006,
reinharz_weighted_2013, hammer_fixed-parameter_2018}. Respecting
positive design directly in the sampling procedure was first
introduced by \citet{andronescu_new_2004} and implemented in
\prog{RNA-SSD}. The authors enhanced the \prog{RNAinverse} algorithm
by sampling the initial seed sequence concerning a probabilistic model
to favor low energy target structures. Subsequently,
\citet{busch_info-rnafast_2006} developed a dynamic programming
approach for the seed generation in \prog{INFO-RNA} to find the
sequence that adopts the target structure with lowest energy possible.
The \prog{IncaRNAtion} approach \cite{reinharz_weighted_2013} extended
this to achieve Boltzmann weighted sampling, where a partition function over
all sequences given the target structure is calculated. This allowed
global weighted sampling, where sequences with a stable target
structure are more likely to be drawn from the solution space than
others and thus a bias toward the desired solutions is gained. By
introducing a weighting term to adjust the nucleotide content, they
got rid of previously observed GC-biases
\cite{reinharz_weighted_2013}. However, all these approaches were only
capable of generating single-stable molecules. Only recently,
\prog{RNAredprint} \cite{hammer_fixed-parameter_2018} introduced a
sampling algorithm which basically combines and extends the methods of
\prog{IncaRNAtion} and \prog{RNAblueprint}
\cite{hammer_rnablueprint_2017}. Additionally, \prog{RNAredprint}
allows to include many of the previously mentioned constraints in an
computationally efficient way and thus renders sequence sampling more
powerful.

\begin{sloppypar}
A different approach to gain similar functionality was introduced with
\prog{RNAiFold}~\cite{garcia-martin_rnaifold:_2012} and its
multi-state design successor \cite{garcia-martin_rnaifold_2015}. The
used constraint programming framework allows to define the desired
constraints in a convenient way in terms of programming. These include
structural constraints as well as sequence soft-constraints such as
compatibility to amino acid sequences. They can be specified either
exactly or based on \prog{BLOSUM62}
similarity\footnote{http://bioinformatics.bc.edu/clotelab/RNAiFold}.
However, the generally applicable constraint programming framework has
the disadvantage of having limited run-time performance, as the
constraint dependencies are getting more complex. Nevertheless,
\prog{RNAifold} allows to either enumerate all sequences of the
solution space or, with the extension of the Large Neighborhood Search
(LNS), to explore unreasonable big solution spaces partially.
\prog{RNAredprint} as well as \prog{RNAifold} are able to count the
number of possible solutions and thus report if no sequence exists for
the chosen inputs.
\end{sloppypar}

\subsubsection{Optimization approach for finding desired solutions}
By constraining and weighting the sequences during sampling, design
goals like the presence of specific sequence patterns, structure
compatibility or even positive design are achievable. However, more
complex goals cannot be directly accomplished yet. A prominent example
is negative design, which aims to get rid of competing states in the
ensemble of structures to achieve high occupancy of the target
state(s). Thus, \ac{RNA} design is often described as an combinatorial
optimization problem, which can be solved by iteratively tackling the
forward problem, \ie \ac{RNA} structure prediction. Such approaches
can be generally dissected into a sequence generation method
(\autoref{sub:sequencegen}), an objective function to evaluate 
the generated sequences (\autoref{sub:designgoals}), and an
optimization strategy which decides whether to keep or reject a
generated solution, which is discussed in the following.

The first design programs mainly used the adaptive walk optimization,
where a new sequence variant is only accepted if it has a better
objective score than the currently best candidate
\cite{hofacker_fast_1994, flamm_design_2001}.
\citet{zadeh_nupack:_2011} included a rejection list of unfavorable
mutations in \prog{NUPACK:Design} to save computation time.
\prog{INFO-RNA} and \prog{RNA-SSD} also accept worse scoring sequences
with a fixed probability to be able to escape local minima
\cite{andronescu_new_2004, busch_info-rnafast_2006}. Simulated
annealing, where this acceptance probability decreases during the
optimization run, was implemented in \prog{RNAexinv},
\prog{ARDesigner} and \prog{RiboMaker} \cite{shu_ardesigner:_2010,
rodrigo_ribomaker:_2014, avihoo_rnaexinv:_2011}.

\citet{kleinkauf_antarna_2015} developed \prog{antRNA}, a nature
inspired ant colony optimization algorithm. Here, ants traverse a
decision tree which lists all allowed nucleotides for each position of
the \ac{RNA} sequence and leave a pheromone trail depending on the
quality of the overall solution. Successive ants are then weighting
their decisions by the pheromone amount along the trail.

Other nature inspired methods follow principles of evolution theory.
A population of sequences is evolved by applying mutation and
recombination to generate new offsprings, which are subsequently
evaluated and selected by their overall fitness. \prog{Frnakenstein}
\cite{lyngso_frnakenstein_2012} implements such a optimization
approach and in \prog{MODENA} a multi-objective genetic algorithm
based on these principles is applied \cite{taneda_modena:_2010,
taneda_multi-objective_2015}. Only recently,
\citet{Rubio-LargoMultiobjective:2018} also used evolutionary
computation and a multi-objective strategy to optimize for standard
design goals. \citet{ramlan_design_2011} even developed multiple
versions of their design tool with different optimization techniques,
including a sorting-bins multi-objective optimization \prog{multiSrch}
and a non-deterministic stochastic variant \prog{StochSrchMulti}.

Although various optimization strategies were applied for the \ac{RNA}
design problem, there exists to our knowledge no study that compares
the efficiency or applicability of the different methods in equivalent
context. The latter would require to use the same sampling method,
objective function, and inputs and to vary only the applied optimization
algorithm to observe the differences in terms of run-time and quality.
However, from the little knowledge about the characteristics of the solution
landscapes, it seems that very simple optimization strategies such as gradient walks
are already sufficient. More insights into the
solution landscapes would help to define a proper neighborhood
relationship that connects \ac{RNA} sequences with similar properties. This
could help to find an efficient way to traverse the solution space. \citet{hammer_rnablueprint_2017} tested the performance
of optimization runs using different neighborhood relations and step
sizes. Interestingly, a random mixture of small and large step-sizes
performed best on a selected set of design instances. We know that
properties of individual sequences are closely connected to the
secondary structure and that the relationship between sequence and
structure is quite complex. Thus, it might be possible to enhance
optimization methods by pursuing the studies on sequence structure
maps, neutral networks and shape space covering
\cite{schuster_sequences_1994, gruner_analysis_1996,
gruner_analysis_1996-1, reidys_generic_1997} with respect to \ac{RNA}
design criteria.

\subsubsection{Filtering and \textit{in silico} analysis}
Ideally, any design goal can be achieved by solving an optimization
problem, which means that all generated candidates would perfectly
fulfill the given requirements. However, as this is not the case, a
subsequent \textit{in silico} analysis step is often needed for
various reasons. For example, some studies did not use an optimization
approach at all and instead enumerated all their candidates. Thus,
they rely on such an analysis step to select solutions by ranking and
filtering with respect to certain criteria and threshold boundaries
\cite{wachsmuth_novo_2013, Domin:2016}. Also, according to
\citet{garcia-martin_rnaifold:_2012}, an exhaustive quality
determination is essential after the candidate enumeration step of the
applied constraint programming approach.

A big advantage of this step is the possibility to sort and filter the
candidates with respect to secondary design goals, which could not be
included in the main optimization approach. These goals might be
computationally too demanding to be iteratively calculated for each
candidate sequence during the optimization. Examples are genome wide
off-target searches for designed trans-\acp{RNA} or exhaustive kinetic
folding experiments \cite{badelt_chapter_2015,
findeis_silico_methods_2018}. By ranking the solutions accordingly, it
is possible to emphasize parts of the design objectives and choose
candidates which fulfill various goals best, an approach which is
similar to solving a posteriori multi-objective problem
\cite{deb_multi-objective_2014}. There, many Pareto optimal solutions
are generated and the user can choose a posteriori which goal is
important and thus heavily weighted. Another reason for the necessity 
of an in silico analysis step could be that
a careful specification of design
goals and constraints was missing and thus an enormous list of
computationally generated candidates has been produced. A ranking and
filtering is then the only option to reduce the amount of candidates
to a number that can be handled in the laboratories.\medskip

Many of the mentioned tools deliver a complete package to convert the
desired features of the novel \ac{RNA} molecule into resulting
sequences by doing many complex tasks on the way. In this section, we
dissected these programs, collected and classified the underlying
methods, and highlighted the algorithmic advances recently made. As
design problems are usually very diverse, delivering a complete
software package with specific design goals, sampling methods and
optimization strategies is not very useful for biologically relevant
applications. Thus, we think that computational tools for rational
\textit{de novo} design only succeed if they are built as software
components which can be flexibly combined to solve a specific design
task \cite{hammer_rnablueprint_2017, findeis_silico_methods_2018}. It
should at least be possible to adapt the objectives, constraints and
other prerequisites of the tools to real world scenarios in order to
serve biologically meaningful applications.

\section{Experimental analysis of the generated candidates}
\label{sec:expanalysis}
The main goal of the previous computational design was to connect the
characterized components and add novel functionality with respect to
the biological target environment. With functional testing assays, we
can investigate whether the candidates exhibit proper overall
performance. However, the causes for frequent discrepancies between
computational predictions and the biological testing results are
manifold. Important influences of the target environment could have
been missed in the model, utilized building blocks might have been
disrupted due to the new context, or the model failed to describe the
new mechanism. Detailed biological and biochemical characterization of
the novel \ac{RNA} device is required to be able to distinguish these
aspects and to react by fine tuning parameters of the computational
design or by identifying and including missing design objectives.

\subsection{Functional testing}
The experimental strategy for the initial functional evaluation
strongly depends on the basic regulatory mechanism of the device to be
tested. In general, \ac{RNA}-based regulators either adopt alternative
structures in order to present or mask certain regulatory elements
(riboswitches) or to fold into an active or inactive ribozyme-based
endonuclease (aptazymes) \cite{Hallberg_invivo_2017, Etzel:2017}. A
further example is to use trans-acting \acp{RNA} that specifically
bind to a target transcript and thereby regulate expression. In the
following, we briefly summarize experimental approaches for the
functional testing of such \ac{RNA} devices.

\paragraph{Testing at the protein level}
Synthetic devices --- riboswitches, aptazymes, trans-acting \acp{RNA}
and others --- are usually designed to regulate expression of a
certain target gene. Hence, their overall functionality can be tested
using a standard reporter like \ac{GFP}, LacZ,
auxotrophic or antibiotic resistance markers. While these genes are
easy to handle and give reliable read-outs to measure the response
ratio of the system, they have certain disadvantages to be considered.
Analysis of \ac{GFP} expression or other fluorescent proteins is rapid
and straightforward, as it can be monitored without disrupting the
cells and even allows for qualitative and quantitative \ac{FACS}. A
LacZ analysis is more labor-intensive, as suitable cell
extracts have to be prepared. If the corresponding enzyme,
$\beta$-galactosidase, is expressed, it catalyzes cleavage of ortho-nitrophenyl-$\beta$-galactoside into galactose and ortho-nitrophenol, a yellow reaction product that is easy to quantify.
The
complex analytic method and considerable fluctuations of the
$\beta$-galactosidase enzymatic activity makes \ac{GFP} --- despite a
longer protein folding time --- more suitable for functional testing.
However, both reporter are perfectly suited for quantification
\cite{Soboleski_GFP_2005, Jensen_bgaB_2017}. Auxotrophic and
resistance markers can be tested in replica-plating, allowing for
rapid selection of positive/functional constructs but a quantitative
analysis of these markers is difficult. A very different read-out
system was used by the Gallivan lab, where functional riboswitch
constructs regulate the expression of CheZ, a protein involved in
chemotaxis. As a result, active constructs could be identified by the
migration of the bacterial host cells towards the source of the ligand
molecule \cite{Topp_GuidingBacteria_2007}.

\paragraph{Testing at the \ac{RNA} level}
In most approaches, riboregulators are designed to control either
transcription or translation of a certain target \ac{RNA}. Yet, it is
possible that the \ac{RNA} construct follows a different regulatory
principle than expected \cite{Fowler_FACS_2008}. Accordingly, it is
important to identify whether such devices act as intended. To
identify regulation at the \ac{RNA} level, several approaches are
possible. One of the most obvious strategies is Northern blot, where
specific oligonucleotide or antisense \ac{RNA} probes identify the
expressed \ac{RNA}. Depending on the experimental setup, this analysis
provides further important information like expression level and ---
for instance in the case of aptazyme-mediated cleavage --- size and
abundance of individual \ac{RNA} fragments. Similarly, \ac{qRT-PCR}
can be used to investigate \ac{RNA} expression and its turnover. A
less labor-intensive approach is the usage of light-up \ac{RNA}
aptamer sequences as reporter. Here, \ac{SELEX}-derived aptamers like
Spinach, Broccoli or Mango can be used that represent functional mimicries of fluorescent proteins~\cite{Ouellet:2016}. A disadvantage of light-up \ac{RNA}
aptamers is their rather low sensitivity. To overcome this problem,
serial arrangements of multiple aptamer copies are used
\cite{Filonov.2016, Zhang.2015}. Furthermore, the presence of high
backgrounds that are not observed \textit{in vitro} necessitate
proper control experiments and complicate the general \textit{in vivo}
applicability \cite{Ilgu:2016}. As an alternative, the J\"aschke lab
developed turn-on aptamers that either bind a fluorophore (like
sulforhodamine B) or a quencher molecule conjugated to a fluorophore.
If these ligands interact with the aptamer, the fluorescence signal
strongly increases and can be measured without disrupting the host
cell \cite{Arora_Quencher_2015, Sunbul_RNAviz_2018}. However, the
general applicability as a read-out system for synthetic \ac{RNA}
regulators still needs to be shown.

\paragraph{Upscaling of functional testing}
The described read-out systems for functional \ac{RNA}-based
regulators are rather limited in terms of the number of samples to be
tested. Computational design approaches can produce hundreds of putative
equally good candidate sequences and valuable information about their
prediction performance can only be gained if many of them are
evaluated experimentally. Hence, it is desirable to test not only a
few candidates but to scale up functional testing to several hundred
or more. \ac{FACS} is an ideal strategy to do so
\cite{lynch_throughput_2007, Lynch_FACS_2009}. However, the
applicability of this approach is currently limited to fluorescence
read-out on the protein level, because usage of the available light-up
\ac{RNA} aptamers suffers from the low sensitivity, as described
above. If microfluidics systems such as the one applied to optimize
the Spinach aptamer \cite{Ryckelynck_Microfluidics_2015} are
applicable as \textit{in vitro} proxy needs to be tested.

\subsection{Characterization of individual constructs}
Independent of the outcome of the functional testing, it is important
to understand and verify the mechanistic basis of the regulation in
order to improve the computational design. Thus, it is mandatory to
confirm and quantify the accomplishment of the initially specified
design goals individually, which includes to assess the functionality
of any utilized building block in the new context, to verify the
mechanism of newly added functionality and to check for missing
indispensable design goals. Especially the latter is a frequent cause
of dysfunctional designs, as there are currently many uncertainties in
the predictions, which include inaccurate or missing model parameters
as well as unconsidered influences of the target environment.
Frequently missing aspects are degradation, RNases, \ac{RNA} binding
proteins and chaperones, off-target interactions, tertiary interactions
and kinetic or co-transcriptional effects. In the following, we give a
short --- and definitely incomplete --- overview about currently used
investigations.

\paragraph{Structure analysis of \ac{RNA} devices}
In the case of riboswitches or aptazymes, the regulatory principle is
based on the (predicted) structures and their ligand-dependent
rearrangement. To investigate whether the intended structures are
indeed dominating the structural ensemble in the target environment,
several approaches are feasible. The most straightforward --- but
restricted to \textit{in vitro} investigation --- is in-line probing.
End-labeled \textit{in vitro} transcripts are incubated in a buffered
solution over a long period of time, allowing that unpaired
nucleotides in single-stranded regions adopt a conformation where the
2'OH group is in line with the phosphorus atom and the adjacent oxygen
of the neighboring phosphodiester bond. This in-line configuration
allows for a nucleophilic attack of the 2'hydroxyl, resulting in a
specific cleavage of the single-stranded site
\cite{Regulski_inlineprobing_2008}. Based on the resulting band
pattern in polyacrylamide gel electrophoresis, the structural
organization of such transcripts can be identified. Similarly,
lead-induced cleavage can be used to generate band patterns indicative
of single-stranded elements, where the hydrated Pb\textsuperscript{2+}
acts as a Br\"onsted base and abstracts the proton from the 2'OH in
the ribose, rendering it highly nucleophilic. An alternative is SHAPE,
where single-stranded nucleotides are modified by chemical treatment
(DMS and NMIA as examples) and detected by \ac{RT} stops at these
positions \cite{Wilkinson.2006, Deigan.2009}. While in-line probing
can only be used \textit{in vitro}, lead probing was successfully
applied \textit{in vivo} for individual transcripts, where the fragile
positions were identified based on primer extensions in reverse
transcription. Recently, also structure probing methods for
high-throughput \textit{in vivo} analyses were established. Methods
like SHAPE-Seq, DMS-Seq, PARS, FragSeq, MOHCA \cite{Lucks.2011,
Kertesz.2010, Mortimer.2012, Rouskin.2014, Cheng.2015} combine the
sensitivity of single stranded \ac{RNA} for cleavage or modification
with massive parallel sequence analysis. While such approaches are
more suited for whole structurome investigations, they can in
principle also be used to determine the structural composition of
individual \ac{RNA} constructs.

An unavoidable fact for a switching \ac{RNA} molecule is the side by
side existence of multiple structures in the ensemble, which
inevitably results in overlapping signals when structurally analyzed.
Yet, based on SHAPE mapping data, a first approach to identify the
individual conformations was recently developed
\cite{Spasic_NMR_2018}, indicating that it might be possible in the
near future to investigate \ac{RNA} devices with multiple structures
at equilibrium.

\paragraph{Characterization of ligand interaction and recognition}
In the case of aptamer-dependent constructs, it is valuable to
identify the actual affinity and interaction with its trigger molecule
in the actual target environment. This includes experiments to verify
the presence of the ligand in the actual reaction compartment, \eg the
bacterial cell. Here, mostly \textit{in
vitro} investigations are performed, but \textit{in vivo} approaches
might also be applicable. Again, a rather easy approach is in-line
probing in the absence and presence of the ligand
\cite{Regulski_inlineprobing_2008}. Originally developed to determine
the structure of the ligand-bound \ac{RNA} and its affinity in term of
K\textsubscript{d}, this method can be equally used for riboswitches
\cite{Kim.2007}. Here, also lead probing as well as \textit{in vitro}
SHAPE should be applicable. Other approaches like \ac{SPR}, \ac{EMSA}
or \ac{MST} might also be usable. However, if the interacting ligand
is a rather small molecule (as in most cases), the impact on the
mobility of the transcript might be too small to be reliably detected.

\paragraph{Determine \ac{RNA} stability and degradation}
\ac{RNA} stability and degradation are important factors when it comes
to real-world applications of artificial \ac{RNA} devices. As they are
often ignored in current design models, detailed information about the
stability and the degradation over time is essential in order to
detect undesired effects. Previously described \ac{RNA} quantification
experiments can also be applied here, such as lifetime experiments
with Northern blot or \ac{qRT-PCR}. These methods are commonly used to
investigate and measure the \ac{RNA} decay.

\paragraph{Characterization of the transcript ends}
Previously mentioned experiments are not suited to verify that the
complete \ac{RNA} molecule, from its intended start nucleotide to the
end, is present and appropriate. However, the actually occurring 5'-
and 3'-ends of the designed \ac{RNA} are of specific interest, as they
give information about, for example, the transcription start site,
termination site or cleavage positions in an aptazyme. As
transcription start sites can differ from promoter to promoter, it is
important to identify the exact sequence composition, since even a
single unexpected nucleotide can lead to misfoldings and
loss/reduction of function. The method of choice is 5'- (or 3'-)
\ac{RACE}, where in the basic version of the protocol oligonucleotide
adapters are ligated to the 5'- (or 3'-) end of the transcript
followed by cDNA synthesis. The resulting product is PCR-amplified by
the usage of a target \ac{RNA}-specific primer. The adapter-ligation
site (corresponding to the \ac{RNA} end) is then identified by
sequencing. Depending on the actual problem, many different \ac{RACE}
versions are available for \ac{RNA} end investigation in prokaryotes
as well as eukaryotes \cite{ScottoLavino.2006, Machida.2014,
Tillett.2000}.

\paragraph{Analysis of kinetic effects}
The actual functionality of multi-state \ac{RNA} devices often depend
on precisely timed kinetic effects, which are typically difficult to
comprise in folding predictions. For example, an intrinsic terminator
has to be present at the exact right time during the transcription
process in order to efficiently stop the \ac{RNA} polymerase. A
fortiori, experimental examination of intermediate structural
conformations during transcription is necessary to understand the
function of many \ac{RNA} devices, especially of riboswitches.
Although \citet{watters_cotranscriptional_2016} managed to obtain
valuable structural folding information of the \textit{crcB} fluoride
riboswitch during transcription by utilizing co-transcriptional
SHAPE-seq and \citet{helmling_NMR_2017} successfully applied NMR
spectroscopy to investigate co-transcriptional intermediate structures
of the I-A type 2'dG-sensing riboswitch from \textit{Mesoplasma florum}, such
experiments are tedious and labor intensive and thus of limited use
for design applications. Moreover, for a proper verification of the
ligand-binding model, not only the binding affinity but also the speed
of the reaction --- the binding rate --- is a crucial factor. Therefore, \citet{Schaffer_Ligand_2014} applied
methods like in-line probing and \ac{SPR} to investigate the kinetics
and thermodynamics of the ligand to AdoCbl–Riboswitch interaction.
Alternative fluorescent ligands with similar binding behavior and
fluorescent nucleobases can be also used to obtain association rates
by stopped-flow fluorescence \cite{Xu_nucleobase_2017,
Gilbert_fluorescent_nucleobase_2006}.
\medskip

Nonetheless, for the purpose of a appropriate cost-benefit ratio, many
of these characterization experiments are often spared. If so, at
least functional testing assays with smartly designed control
constructs to derive valuable characteristic details of the
constructed \ac{RNA} molecules should be performed. For example,
proper controls for a transcriptional riboswitch could be a construct
which comprises only the terminator or only the aptamer and the
\ac{RBS} with the otherwise exact same sequence as the designed
construct. Also well-considered time series of standard \ac{RNA} or
protein read-out experiments can deliver important details about
intermediates and kinetic effects. Furthermore, if it is not possible
to derive worthwhile information from the characterization
experiments, randomization and a subsequent screening or selection can
of course always be applied to optimize the construct. However, it is important to
investigate the optimized constructs and gain valuable knowledge about
their (improved) functionality in order to enhance the design model
for future applications.

\section{Conclusion}
Rational \textit{de novo} design always requires some initial
knowledge about the functional mechanism, the characteristics of
components and the target environment. For a riboswitch regulating at
the transcriptional level, for example, the most obvious needed
information is termination efficiency and mechanics of \ac{RNA}
structural refolding. However, other information is also important,
\eg ribosomal binding efficiency, which determines translation
initiation for reporter gene expression. The combined information then
makes up the functional model, which is part of the overall design
goal.

Previously, this information was determined only for specific
sequences, which limited the ability to optimize the overall
computational design. The usage of a catalog of sequences for the
individual building blocks will improve this situation to some extent.
Currently, data-driven models are developed to provide enough detail
and variability for this purpose. However, at the moment such models
are rare and rational designs are often built with little knowledge
about the biological environment and the functional mechanism. To
compensate these uncertainties, lots of subsequent laboratory work is
required, such as functional testing and characterization experiments
to verify the functionality of the individual components in the new
context. Due to recent advances, well-designed high-throughput
selection and screening pipelines are able to measure function of many
varying \ac{RNA} sequences, thus producing the amount of standardized
data necessary to built detailed and extensive models
\cite{carlson_elements_2018, Borujeni:2013, Cambray:2013, Chen:2013,
Kosuri:2013}.

We realized that rational \textit{de novo} design was
sometimes performed without much support from \textit{in silico}
calculations. We speculated that the reason might be a lack of trust
in the available \ac{RNA} design tools due to the limitations of
computational prediction. Examples are the inability to consider
pseudoknots, the exclusion of non-Watson-Crick pairings, the often
neglected influence of ions in the predictions and models that do not
accurately reflect ligand-\ac{RNA} interactions. Furthermore,
predictions might miss the structural influences of proteins and
organic molecules in the cell and cannot explain possible structural
differences between \textit{in vitro} and \textit{in vivo}
experiments. 

This uncertainty is increased by the frequent lack of experimental
verification or real-world synthetic biology applications. Those are
often missing as experiments are time-consuming and expensive.
Instead, the superiority of novel algorithms is often concluded with
extensive benchmarks towards some seemingly arbitrary goals with no
obvious biological applicability. We understand that benchmarking
requires standardized inputs in form of an extensive test set and a
precise definition of a perfect solution. These prerequisites are hard
to be fulfilled by laboratory experiments. Nevertheless, in our
opinion, the best benchmark of an \ac{RNA} design tool is to show its
ability to reliably produce functional \ac{RNA} devices in real-world
settings. \prog{NUPACK:Design}, for instance, has been used to
implement so-called ``toehold'' switches \cite{Green:2014}, small
transcription-activating \acp{RNA} \cite{chappell_computational_2017},
small conditional \acp{RNA} to regulate \ac{RNA} interference
\cite{Hochrein:2018} and to control the formation of complex
nanostructures \cite{Geary:2014}. The \prog{RNAiFold} approach has
been adapted and applied to design hammerhead ribozymes
\cite{dotu_complete_2014}, regulate \ac{IRES} activity
\cite{Fernandez-Chamorro:2016} and to realize temperature-sensitive
\ac{IRES} elements \cite{Garcia-Martin:2016}. Thermoswitches have been
designed utilizing \prog{switch.pl} \cite{Waldminghaus:2008}. In most
cases, the developers of the software are among the authors of all
these application studies. This indicates that expert knowledge about
the \textit{in silico} design approach is beneficial for a successful
\textit{in vivo} implementation.

Of course, it is possible to completely skip computational design and
instead utilize high-throughput screening or selection approaches to
obtain complete \ac{RNA} devices with the advantage that no model is
required \cite{Fowler_FACS_2008, harbaugh_screening_2018,
page_engineering_2018}. In such studies, the measurement directly
indicates the functionality of the individual devices from the
variable pool. However, the results of these studies do not generalize
to other design problems. By contrast, if appropriate data can be
collected in experimental design projects, this data could provide a
basis for a computational model that might accelerate future design
projects. \citet{TownshendHighthroughput:2015} use a workflow of \ac{FACS} and
next-generation sequencing to achieve this
\cite{SchmidtSynthetic:2019}.

On the experimental side, new approaches are needed to be able to
measure and verify important parameters of \textit{in silico} models
in an exact and high-throughput manner. This would allow the design
and validation to form a stronger feedback loop in the design process,
because directly measured data will lead to improved models.

Our vision is that it is possible to quickly and consistently design
\acp{RNA} with the desired functionality for their use as artificial
logic in gene-regulation, as scaffolding devices or in
therapeutic applications. A fundamental stumbling block is the
relative lack of effective collaboration between computational and
experimental scientists. We believe that the inadequate level of
communication between these fields is limiting the ability to develop
quantitative mechanistic models. Moreover, individual pipeline steps
are developed independently of one another, and thus fail to function
optimally together. Finally, much research is ineffective in promoting
progress, because it pursues arbitrary goals that are not biologically
relevant.

To address this problem, there is a need to focus on the practical
applicability of research, including novel algorithms, wet lab
techniques and design projects. There is a need for computational
\ac{RNA} design tools that are useful for their actual purpose, experiments that
measure properties needed for \textit{in silico} models and valuable
data generated when performing screening and selection experiments.

\section*{Acknowledgments}
This work has been supported by the German Network for Bioinformatics
Infrastructure (de.NBI) by the German Federal Ministry of Education
and Research (BMBF; support code 031A538B), and by the German Research
Foundation (DFG; grants STA 850/15-2 and MO 634/9-2). The project
RiboNets acknowledges the financial support of the Future and Emerging
Technologies (FET) program within the Seventh Framework Program for
Research of the European Commission, under FET-Open grant number:
323987. We thank Zasha Weinberg for fruitful discussions and comments
on the manuscript.

\section*{Abbreviations}
\begin{acronym}[*****]
 \acro{UTR}[UTR]{untranslated region}
 \acro{mRNA}[mRNA]{messenger RNA}
 \acrodefplural{mRNA}[mRNAs]{messenger RNAs}
 \acro{tRNA}[tRNA]{transfer RNA}
 \acrodefplural{tRNA}[tRNAs]{transfer RNAs}
 \acro{RNA}[RNA]{RiboNucleic Acid}
 \acro{RNAs}[RNAs]{RiboNucleic Acids}
 \acro{SELEX}[SELEX]{Systematic Evolution of Ligands by EXponential enrichment}
 \acro{MFE}[MFE]{minimum free energy}
 \acro{RBS}[RBS]{ribosome binding site}
 \acro{SD}[SD]{Shine Dalgarno}
 \acro{GFP}[GFP]{green fluorescent protein}
 \acro{FACS}[FACS]{fluorescence-activated cell sorting}
 \acro{RACE}[RACE]{rapid amplification of cDNA-ends}
 \acro{TSS}[TSS]{transcription start site}
 \acro{RT-PCR}[RT-PCR]{reverse transcription polymerase chain reaction}
 \acro{SHAPE}[SHAPE]{selective 2'-hydroxyl acylation analyzed by primer extension}
 \acro{qRT-PCR}[qRT-PCR]{quantitative reverse transcription-polymerase chain reaction}
 \acro{cDNA}[cDNA]{complementary DNA}
 \acro{ecoli}[\textit{E. coli}]{\textit{Escherichia coli}}
 \acro{bsubtilis}[\textit{B. subtilis}]{\textit{Bacillus subtilis}}
 \acro{qPCR}[qPCR]{quantitative real-time PCR}
 \acro{RT}[RT]{reverse transcription}
 \acro{SPR}[SPR]{surface plasmon resonance spectroscopy}
 \acro{EMSA}[EMSA]{electrophoretic mobility shift assay}
 \acro{MST}[MST]{microscale thermophoresis}
 \acro{IRES}[IRES]{internal ribosome entry site}
\end{acronym}

\bibliography{publications}

\end{document}